\documentclass[manuscript=article]{achemso}

\usepackage{graphicx}
\usepackage{dcolumn}
\usepackage{bm}
\usepackage{siunitx} 
\usepackage[artemisia]{textgreek}
\usepackage[utf8]{inputenc}
\usepackage{natbib}
\usepackage[T1]{fontenc}
\usepackage{mathptmx}

\author{G. Gruber}
\author{C. Urgell}
\author{A. Tavernarakis}
\author{A. Stavrinadis}
\author{S. Tepsic}
\affiliation{ICFO – The Institute of Photonic Sciences, Av. Carl Friedrich Gauss 3, 08860 Castelldefels (Barcelona), Spain}%
\author{C. Mag\'en}
\affiliation{Instituto de Ciencia de Materiales de Aragón (ICMA), Universidad de Zaragoza-CSIC, 50009 Zaragoza, Spain}
\alsoaffiliation{Laboratorio de Microscopías Avanzadas (LMA), Instituto de Nanociencia de Aragón (INA), Universidad de Zaragoza, 50018 Zaragoza, Spain}%
\author{S. Sangiao}
\affiliation{Instituto de Ciencia de Materiales de Aragón (ICMA), Universidad de Zaragoza-CSIC, 50009 Zaragoza, Spain}
\alsoaffiliation{Laboratorio de Microscopías Avanzadas (LMA), Instituto de Nanociencia de Aragón (INA), Universidad de Zaragoza, 50018 Zaragoza, Spain}%
\author{J. M. de Teresa}
\affiliation{Instituto de Ciencia de Materiales de Aragón (ICMA), Universidad de Zaragoza-CSIC, 50009 Zaragoza, Spain}
\alsoaffiliation{Laboratorio de Microscopías Avanzadas (LMA), Instituto de Nanociencia de Aragón (INA), Universidad de Zaragoza, 50018 Zaragoza, Spain}%
\author{P. Verlot}
\affiliation{School of Physics and Astronomy - The University of Nottingham, University Park, Nottingham NG7 2RD, United Kingdom}%
\author{A. Bachtold}
\email{adrian.bachtold@icfo.eu}
\affiliation{ICFO – The Institute of Photonic Sciences, Av. Carl Friedrich Gauss 3, 08860 Castelldefels (Barcelona), Spain}%

\title{Mass sensing for the advanced fabrication of nanomechanical resonators}

\begin{document}

%
%
%


\begin{abstract}
We report on a nanomechanical engineering method to monitor matter growth in real time via e-beam electromechanical coupling. This method relies on the exceptional mass sensing capabilities of nanomechanical resonators. Focused electron beam induced deposition (FEBID) is employed to selectively grow platinum particles at the free end of singly clamped nanotube cantilevers. The electron beam has two functions: it allows both to grow material on the nanotube and to track in real time the deposited mass by probing the noise-driven mechanical resonance of the nanotube. On the one hand, this detection method is highly effective as it can resolve mass deposition with a resolution in the zeptogram range; on the other hand, this method is simple to use and readily available to a wide range of potential users, since it can be operated in existing commercial FEBID systems without making any modification. The presented method allows to engineer hybrid nanomechanical resonators with precisely tailored functionality. It also appears as a new tool for studying growth dynamics of ultra-thin nanostructures, opening new opportunities for investigating so far out-of-reach physics of FEBID and related methods.
\end{abstract}


\noindent Keywords: Mechanical resonators, NEMS, nanofabrication, mass sensing, carbon nanotube, electron microscopy\\
\\
\noindent Nanomechanical devices are exquisite sensors of mass deposition \cite{bib:YangNL2006, bib:ChiuNL2008, bib:GilSantosNN2010, bib:ChasteNature2012} and external forces.\cite{bib:MaminAPL2001,bib:GavartinNN2012,bib:MoserNatureNanotech2013,bib:deBonisNL2018,bib:HeritierNL2018} These sensing capabilities enabled advances in mass spectrometry,\cite{bib:HanayNN2012,bib:HanayNN2015,bib:DominguezScience2018} surface science,\cite{bib:WangScience2010,bib:YangNL2011,bib:AtalyaPRL2011,bib:TavernarakisPRL2014,bib:RhenPRB2016,bib:SchwenderAPL2018,bib:NouryPRL2019} scanning probe microscopy,\cite{bib:RossiNatureNanotech2017, bib:LepinayNatureNanotech2017} and magnetic resonance imaging.\cite{bib:RugarNature2004,bib:DegenPNAS2009, bib:NicholPRX2013} The highest sensitivity is achieved with carbon nanotube resonators\cite{bib:ChasteNature2012,bib:deBonisNL2018} because of their tiny mass compared to the other operational mechanical resonators. However, a general challenge with such small transducers is to provide them with a physical function, which can be e.g. magnetic, chemical, or optical. Conventional nanofabrication processes, such as electron-beam lithography and reactive-ion etching, are difficult to employ with such small suspended structures without altering their sensing capabilities. Developing new methods to engineer nanoscale resonators with high precision and providing them with a specific functionality is in high demand as it would enable a whole range of new technological and scientific applications.

In this work we report a nanofabrication method enabling ultra-sensitive, versatile functionalization of carbon nanotube resonators\cite{bib:KhivrichNNano2019, bib:BarnardNature2019, bib:UrgellArxiv2019} inside a scanning electron microscope (SEM). Using focused electron beam induced deposition (FEBID),\cite{bib:NishioJVSTB2005, bib:SawayaAPL2006, bib:FriedliAPL2007, bib:BanerjeeNT2009, bib:SangiaoBJN2017} we report the mass-controlled growth of Pt particles on carbon nanotube nanomechanical sensors, enabling their optomechanical funcionalization.\cite{bib:TavernarakisNC2018} The deposited mass is tracked in real time by monitoring frequency changes of the noise-driven oscillations of the nanotube resonator. Measuring the nanomechanical vibrations relies on e-beam electromechanical coupling\cite{bib:TsioutsiosNL2017, bib:PairisPRL2019} and is accomplished using the same electron-beam as that used for FEBID. We demonstrate the high sensitivity and versatility of this method, which enables us to address mass changes over more than six orders of magnitude, with a resolution down to the zg range.

The samples consist of carbon nanotubes grown via chemical vapor deposition on silicon substrates. The nanotubes stick to the surface due to Van der Waals forces. Some nanotubes extend over the substrate edge, forming cantilevers. We used cantilevers with lengths between 1\,\textmu m and 15\,\textmu m and spring constants between $10^{-7}\,\textrm{N/m}$ and $2.6\times10^{-4}\,\textrm{N/m}$ in order to investigate the robustness of our method. \rm

All SEM and FEBID experiments were conducted in a Zeiss Auriga field emission electron microscope equipped with a gas injection system (GIS). The acceleration voltage of the electron beam was $5\,\textrm{kV}$ and the typical beam current was $200\,\textrm{pA}$. The precursor gas was methylcyclopentadienyl(trimethyl)platinum(IV) in order to grow a Pt deposit onto the sample surface when illuminated by the electron beam.\cite{bib:WnukJPCC2009} All the experiments reported below have been completed with the GIS nozzle being placed $\approx500\,\textrm{\textmu m}$ above the substrate.

\begin{figure*}[t]
\includegraphics[width=\linewidth]{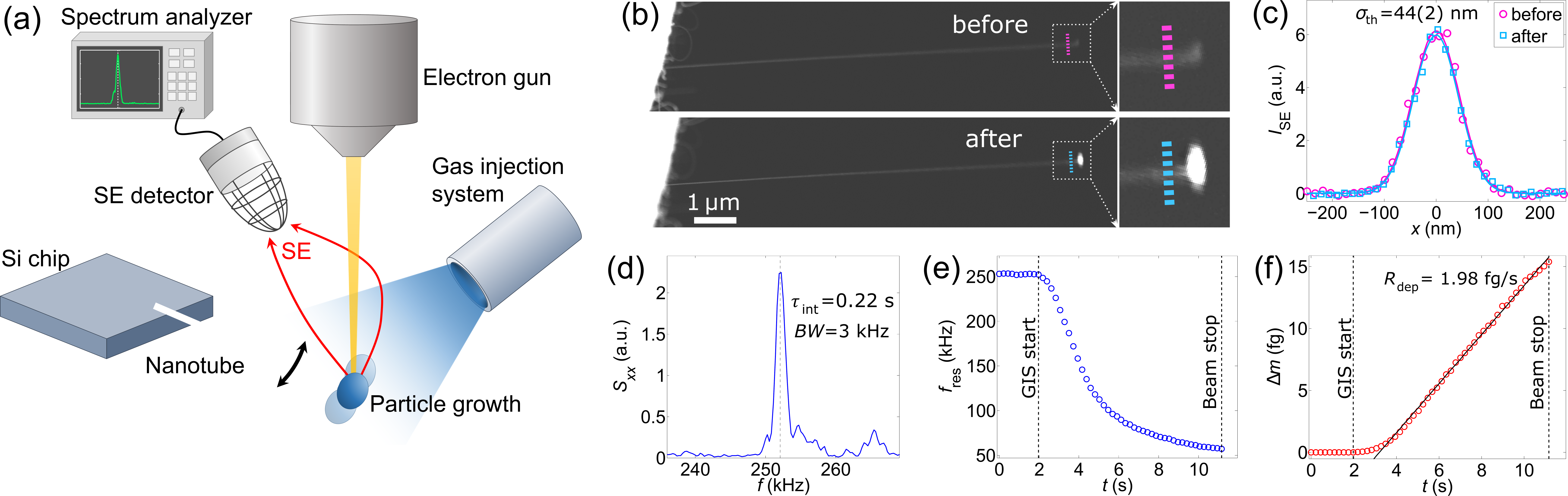}
\caption{(a) Schematic of the setup: The electron beam is set on the apex of the suspended nanotube cantilever, creating a secondary electron (SE) current, which is detected and fed into a spectrum analyzer. Using the gas injection system (GIS) a nanoparticle is grown on the nanotube, resulting in a shift of the observed resonance frequency. (b) SEM images of a nanotube before and after the deposition of a particle, with 3x magnified view of the apex (right side). (c) Profiles of the SE current $I_\textrm{SE}$ along the dashed lines marked in (b) with Gaussian fits (solid lines). (d) Typical resonance signal used to count the resonance frequency. (e) Monitoring of the resonance frequency during the deposition; at $t\approx2\,\textrm{s}$ the GIS valve was opened and at $t\approx11\,\textrm{s}$ it was closed and the beam exposure stopped. (f) Deposited mass determined from (e) using Eq. \ref{eq:deposition}.}
\label{fig:Methods}
\end{figure*}

A schematic of the experimental setup used for the deposition experiments is depicted in Fig. \ref{fig:Methods}(a). The electron beam is set onto the apex of the nanotube in spot mode while monitoring the secondary electron (SE) current $I_\textrm{SE}$. The signal is displayed in the frequency domain via fast Fourier transform (FFT). The data are real-time processed using a fast peak-search custom computer program, enabling us to extract the mechanical resonance frequency at a rate between typically 0.5\,Hz and 5\,Hz.

Figure \ref{fig:Methods}(b) shows a nanotube before and after the deposition process with the deposited particle clearly visible. Furthermore, the free end of the nanotube appears blurred due to the motion fluctuations. The spring constant $k$ can be extracted from the variance of the displacement $\sigma_\textrm{th}^2$ using the equipartition theorem
\begin{equation}
k=\frac{k_\textrm{B}T}{\sigma_\textrm{th}^2}
\label{eq:spring}
\end{equation}
\noindent where $k_\textrm{B}$ is the Boltzmann constant and $T$ is the temperature.\cite{bib:TsioutsiosNL2017} Figure \ref{fig:Methods}(c) shows the SE current profiles taken along the dashed lines marked in Fig. \ref{fig:Methods}(b) before and after the deposition with Gaussian fits to determine $\sigma_\textrm{th}^2$. The resulting spring constant $k=2.1\left(2\right)\times10^{-6}\,\textrm{N/m}$ is the same in both cases. This shows that $k$ is not affected by the deposition process and any permanent changes in the mechanical resonance frequency are consequently associated with mass deposition (see further discussion below). Specific care was dedicated to avoid broadening of the observed peak by back-action phenomena during imaging.\cite{bib:TsioutsiosNL2017} This was achieved by averaging multiple frames using the fastest scanning speed (122\,ms/frame).

The mass of the Pt particle is monitored in real time during its formation. This is done by continuously acquiring the resonance spectrum of the noise-driven vibrations of the nanotube with the electron beam. We typically use high resolution bandwidth settings in order to enable a high sampling rate. Figure \ref{fig:Methods}(d) shows a typically obtained signal used to count the frequency for the mass detection. The resolution bandwidth of the measurement in this case was $BW=3\,\textrm{kHz}$. The resonance frequency $f_\textrm{res}$ relates to the effective mass $m^*$ of the mechanical eigenmode via the equation:
\begin{equation}
f_\textrm{res}=\frac{1}{2\pi}\sqrt{\frac{k}{m^*}}.
\label{eq:resonance}
\end{equation}
\noindent Figure \ref{fig:Methods}(e) shows the evolution of $f_\textrm{res}$ over time. Here, the GIS nozzle was opened at $t\approx2\,\textrm{s}$. The electromechanical interaction then becomes strongly non-linear, resulting in a strong amplification of the electromechanical spectrum and the appearance of a large number of peaks at multiples of the fundamental resonance frequency (not shown here, see Section 1 of Supporting Information). We attribute this behaviour to the increasing interaction volume resulting from the deposition process. Our frequency counting algorithm includes a dynamical discrimination procedure enabling to unambiguously keep track of the fundamental resonance frequency in real-time. As shown on Fig. \ref{fig:Methods}(e), $f_\textrm{res}$ decreases over time, which is the expected evolution in presence of mass adsorption.

The deposition was limited to the apex of the nanotube, such that the spring strength can be reasonably assumed to remain unchanged. Therefore, the deposited mass $\Delta m\left(t\right)$ yields to a frequency shift, independent from the shape of the eigenmode:\cite{bib:HanayNN2012}
\begin{equation}
\Delta m\left(t\right) = \frac{k}{4\pi^2}\left(\frac{1}{f_{\textrm{res},t}^2}-\frac{1}{f_{\textrm{res},0}^2}\right)
\label{eq:deposition}
\end{equation}
\noindent where $f_{\textrm{res},t}$ and $f_{\textrm{res},0}$ are the resonance frequencies measured during the deposition at time $t$ and prior to the deposition, respectively.\cite{bib:YangNL2006, bib:ChiuNL2008, bib:GilSantosNN2010, bib:ChasteNature2012,bib:HanayNN2012,bib:HanayNN2015,bib:DominguezScience2018, bib:WangScience2010,bib:YangNL2011, bib:TavernarakisPRL2014,bib:SchwenderAPL2018} In the limit of high signal-to-noise ratio, the mass determination does weakly depend on the SE emission rate. Additionally, we performed optomechanical measurements\cite{bib:TavernarakisNC2018} in order to gain independent confirmation of the post-deposition mechanical properties (Section 2 of Supporting Information). These measurements ensure that the electromechanical coupling has negligible impact on the mechanical resonance frequency and that the observed changes are due to mass deposition.

Figure \ref{fig:Methods}(f) displays the corresponding evolution of the deposited mass over time. After some transient regime, the deposition becomes linear in time, allowing us to extract the deposition rate $R_\textrm{dep}=1.98\,\textrm{fg/s}$ from a linear fit. At $t\approx11\,\textrm{s}$ the GIS valve was closed and the beam exposure was stopped to avoid spurious growth. The resonance frequency at the end was $f_\textrm{res}=56.1\left(5\right)\,\textrm{kHz}$ and the total mass of the particle seen in Fig. \ref{fig:Methods}(b) is $\left(15.5\pm2.0\right)\,\textrm{fg}$. Optomechanical measurements of this resonator yield to a post-deposition mechanical resonance frequency $f_0=57.04\,\textrm{kHz}$ with a quality factor $Q\approx3000$ at room temperature (Section 2 of Supporting Information). Besides further confirming the mass-induced origin of the measured frequency change, this measurement demonstrates that the deposition using FEBID does not degrade the mechanical properties of the nanotube resonator, which is crucial in the context of functionalizing nanomechanical resonators.

\begin{figure*}[t]
\includegraphics[width=.9\linewidth]{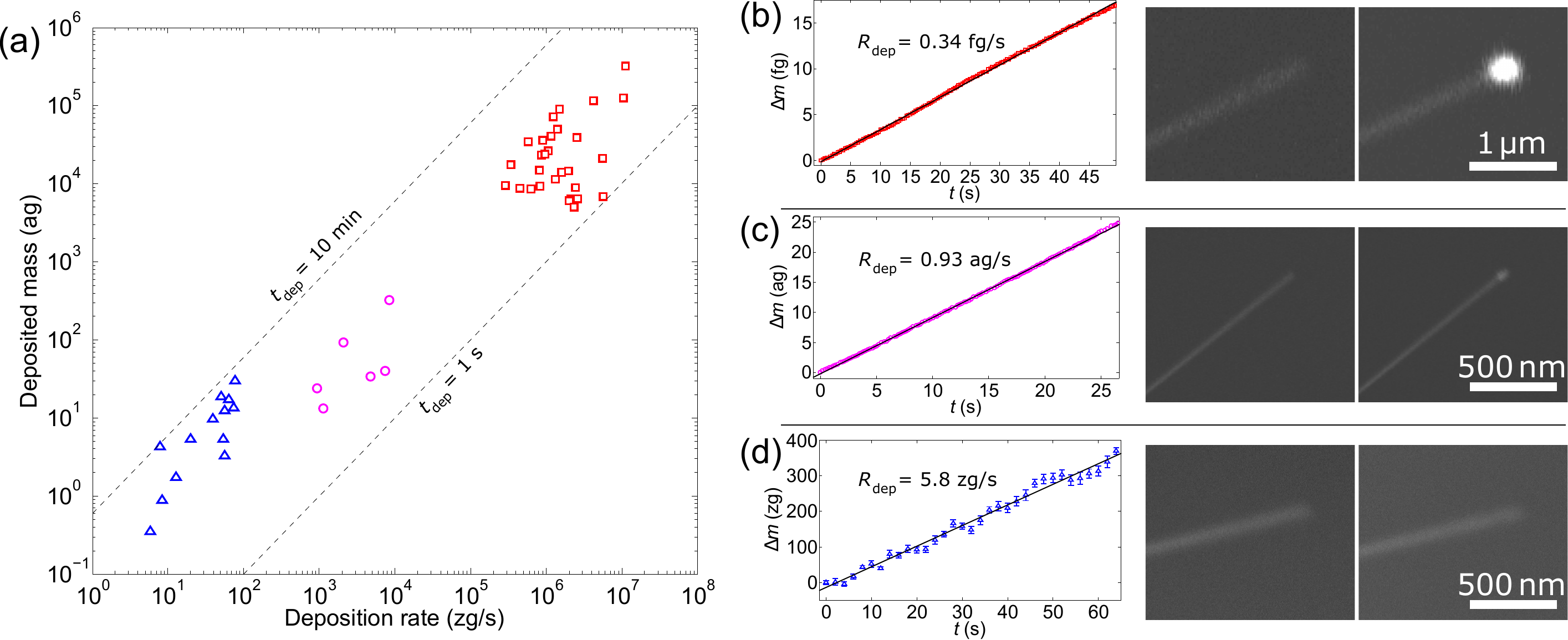}
\caption{(a) Deposition rate and deposited mass for all the fabricated devices, with deposition times $t_\textrm{dep}$ in the range between 1\,s and 10\,min. The different operation modes are marked by different colors, and exemplary measurements are shown in (b)-(d). (b) Mass deposition in default GIS operation mode (GIS nozzle open, precursor in the chamber at a pressure in the range of $p=(7-11)\times10^{-6}\,\textrm{mbar}$). (c) Mass deposition in low-pressure mode (GIS nozzle closed, precursor residuals in the chamber with $p=(1-1.7)\times10^{-6}\,\textrm{mbar}$). (d) Mass deposition in the background vacuum regime (after more than $24\,\textrm{h}$ of pumping, $p=(0.8-1)\times10^{-6}\,\textrm{mbar}$). The SEM images on the right show each nanotube before and after the deposition. The spring constants determined before and after the deposition are $k=6.2\left(5\right)\times10^{-7}$\,N/m for (b), $k=1.57\left(7\right)\times10^{-5}$\,N/m for (c), and $k=1.00\left(3\right)\times10^{-6}$\,N/m for (d).}
\label{fig:Results}
\end{figure*}

Using the above-described methodology, a large set of hybrid nanotube cantilevers were fabricated and characterized. Figure \ref{fig:Results}(a) shows the determined deposition rates and final masses of the deposited particles for each experiment. The dashed lines indicate how the deposition rate and deposited mass are related via the deposition time. The observed variations arise from different modes of operations (see further discussion below), to which have been assigned distinct colors. Note that even within the same mode of operation, the obtained results are widely dispersed. This is because FEBID is a highly complex process where various interdependent parameters may affect the growth rate.\cite{bib:WinklerANM2018} These include the focus of the electron beam, the temperature of the substrate, the temperature and flux of the precursor molecules, and the pressure of residual gas in the chamber. The deposition rate is also affected by the amplitude of the nanotube vibrations, since the amplitude can be larger than the electron-beam diameter, resulting in a net decrease of the effective deposition cross-section. The different GIS operation modes as well as illustrative results are discussed in the following.

We start with the default operation mode of the GIS, which was also used for the measurements in Fig. \ref{fig:Methods}. When the nozzle is opened the precursor gas is released into the chamber resulting in a strong increase of the chamber pressure. The pressure typically saturates in the range $p=(7-11)\times10^{-6}\,\textrm{mbar}$, while the background vacuum pressure is typically $\approx1\times10^{-6}\,\textrm{mbar}$. It results in measured deposition rates between $0.28\,\textrm{fg/s}$ and $11\,\textrm{fg/s}$. Figure \ref{fig:Results}(b) shows a typical measurement in this operation mode, demonstrating a constant deposition rate $R_\textrm{dep}=0.34\,\textrm{fg/s}$ over a time as long as $50\,\textrm{s}$. The deposited mass is more than 30 times larger than the initially measured mass of the nanotube cantilever.

We explored lower Pt deposition rates by reducing the pressure. This is achieved by first purging the GIS nozzle with precursor molecules and then pumping the chamber for several minutes. As such, we investigated deposition of precursor molecules in a pressure range $p=(1-1.7)\times10^{-6}\,\textrm{mbar}$ resulting in observed deposition rates ranging between $0.93\,\textrm{ag/s}$ and $8.5\,\textrm{ag/s}$. A typical mass deposition measurement in this low-pressure regime is displayed in Fig. \ref{fig:Results}(c). The SEM image after the deposition reveals a small Pt particle. The deposition rate $R_\textrm{dep}=0.93\,\textrm{ag/s}$ is equivalent to roughly 2900 Pt atoms or 1800 precursor molecules per second.

The lowest deposition rates were attained by pumping the chamber for more than $24\,\textrm{h}$ with the GIS nozzle closed and heated so residual precursor molecules could desorb from the nozzle and be pumped away. It is assumed that in this regime the chamber gas is predominantly composed of organic molecules resulting in e-beam deposition of amorphous carbon. The base pressure in this background {vacuum} regime was in the range $p=(0.8-1)\times10^{-6}\,\textrm{mbar}$ and the observed deposition rates were between $5.8\,\textrm{zg/s}$ and $77\,\textrm{zg/s}$. The lowest value $R_\textrm{dep}=5.8\,\textrm{zg/s}$ with 2\,s integration time was observed in the experiment shown in Fig. \ref{fig:Results}(d) and is equivalent to about 290 C atoms per second. Computing the Allan deviation of the resonance frequency with 2\,s integration time results in an effective mass resolution of 13\,zg. This estimation includes the spurious contribution of the deposition of C atoms, so that it represents an upper bound of the mass resolution of the nanotube resonator. The deposited mass of $330\;\textrm{zg}$ does not result in a distinctive feature on the nanotube in the SEM images. In this case, the electromechanical measurement enables us to reveal the evolution of the structure that is totally invisible in the SEM image. Besides controlling the growth process, this demonstrates the relevance of e-beam electromechanical coupling as a powerful complementary embedded tool to scanning electron microscopy.

We assessed the material density of a Pt particle and its chemical composition by carrying out scanning transmission electron microscopy (STEM) measurements. The experiments were performed using a $C_\textrm{s}$-corrected FEI Titan transmission electron microscope equipped with a FEI X-FEG high brightness Schottky emitter. The acceleration voltage was $80\,\textrm{kV}$. Chemical analysis was conducted via energy-dispersive x-ray spectroscopy (EDXS) using an EDAX detector. Thickness measurements were performed by electron energy loss spectroscopy (EELS) employing a Gatan Tridiem 866 ERS energy filter.

\begin{figure*}[t]
\includegraphics[width=.8\linewidth]{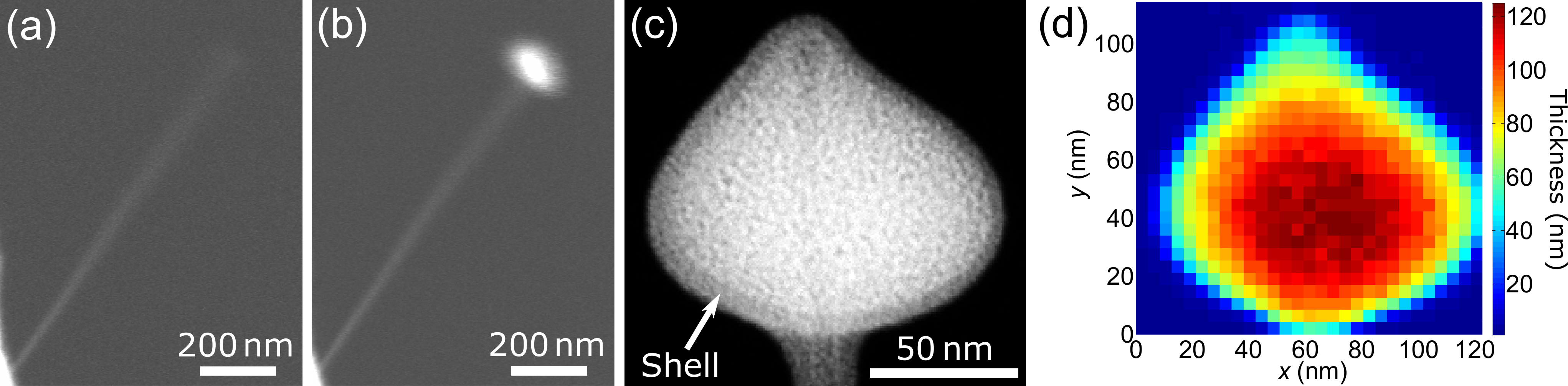}
\caption{(a) and (b) SEM images of a nanotube before and after the deposition of a particle with a mass $m_\textrm{dep}=1.33\,\textrm{fg}$ determined by the resonance frequency measurement. (c) High-angle annular dark-field (HAADF) STEM image of the particle. The visible darker shell is likely the result of the subsequent manipulation of the nanotube with the electron beam (see text). (d) Thickness map of the particle determined by low-loss EELS using the elemental composition of Table \ref{tbl:EDX} and the log-ratio method.\cite{bib:MalisJEMT1988}}
\label{fig:TEM}
\end{figure*}

A nanotube grown on a STEM copper grid is shown in Figs. \ref{fig:TEM}(a) and (b) before and after depositing a mass of $1.33\,\textrm{fg}$. In a subsequent deposition step the nanotube was coated with material down to the clamping point to minimize the effects of motion fluctuations. During this step we avoided exposing the particle directly to the electron beam. Figure \ref{fig:TEM}(c) displays a high-angle annular dark-field (HAADF) image recorded in the STEM and reveals a core-shell structure. The shell appears darker than the core, suggesting that it has a lower density or a lower relative Pt content than the core particle. It is likely that the shell was formed during the second deposition step and the growth induced by secondary electrons. Previous FEBID works showed that Pt atoms assemble together to form nanometer-scale clusters inside an amorphous C matrix.\cite{bib:WnukJPCC2009}

\begin{table}
\caption{Atomic fraction and mass fraction of C, Pt, and O determined by EDXS measurements of the particle in Fig. \ref{fig:TEM}(c).}
\label{tbl:EDX}
\begin{tabular}{lrr}
\hline
Element & Atomic fraction $\left(\%\right)$ & Mass fraction $\left(\%\right)$\\
\hline
Carbon & 84.6 & 41.5\\
Oxygen & 8.8 & 5.8\\
Platinum & 6.6 & 52.7\\
\hline
\end{tabular}
\end{table}

The particle composition was determined by EDXS measurements (Table \ref{tbl:EDX}). The observed oxygen content is attributed to the air molecules that diffuse into the particle during the transfer of the device from the SEM to the STEM. The atomic C:Pt ratio determined by EDXS is 12.8:1. This ratio is somewhat larger than the value 8:1 reported previously,\cite{bib:WnukJPCC2009} suggesting additional amorphous carbon deposition in our experiment, especially during the second deposition step.

Next, we conducted spatially-resolved low-loss EELS measurements to map the particle thickness (Fig. \ref{fig:TEM}(d)). The thickness at each point of the map was determined via the log-ratio method using the electron inelastic mean free path determined by the elemental composition in Table \ref{tbl:EDX}.\cite{bib:MalisJEMT1988} We obtain the volume $V=5.45\times10^{-16}\,\textrm{cm}^3$ for the particle core by integrating the thickness over the map surface, and by subtracting the volume of the 6\,nm thick shell. This results in the density $\rho\simeq2.44\,\textrm{g/cm}^3$ for the particle core using the mass $1.33\,\textrm{fg}$. With this result we are able to estimate the density of the amorphous C-matrix $\rho_\textrm{C}=\left(\rho-\xi\rho_\textrm{Pt}\right)/\left(1-\xi\right)$ where $\xi$ is the normalized atomic Pt concentration in the pseudo-binary composite $\textrm{Pt}_\xi\textrm{C}_{1-\xi}$ and $\rho_\textrm{Pt}=21.45\,\textrm{g/cm}^3$ is the bulk density of Pt.\cite{bib:UtkeAPL2006, bib:FriedliNT2009} Our estimation $\rho_\textrm{C}\simeq0.95\,\textrm{g/cm}^3$ compares well with \it low-quality \rm amorphous (hydro-)carbon deposits, which are typically in the range $(0.3-1.5)\,\textrm{g/cm}^3$.\cite{bib:NishioJVSTB2005, bib:SawayaAPL2006, bib:FriedliNT2009, bib:FriedliAPL2007, bib:RobertsonMSER2002}

The mass monitoring method opens new possibilities to study the growth of ultra-thin nanostructures using FEBID.\cite{bib:FriedliNT2009, bib:FernandezPRB2009, bib:PabloANM2018, bib:BanerjeeNT2009, bib:WnukJPCC2009, bib:ArnoldAFM2018} It may be applied to study how the mass deposition and the material composition depend on experimental growth parameters, such as the electron beam current, the gas-injection rate, and the precursor and substrate temperature.\cite{bib:WinklerANM2018} Our technique is particularly attractive to investigate transients at the beginning of the growth. The good time resolution in the monitoring of the growth rate could be used to test different growth models, e.g. involving various precursor dissociation mechanisms (triggered by primary and secondary electrons), precursor coverage, or thermal effects. It may also be employed to test new precursors and to monitor purification steps aiming to improve the material quality. Furthermore, mass monitoring using our method could be applied to study the growth and the milling with a focused ion beam.

In summary, we have reported a method allowing high-resolution mass monitoring of the growth of a Pt nanoparticle on a nanotube resonator via \it in situ \rm electromechanical readout in a FEBID system. The method can be readily employed in any existing SEM or STEM setup without requiring any further modification. The demonstrated mass and time resolution offers a precise control on the deposited mass to engineer nanomechanical sensors, especially since various materials can be grown with FEBID.\cite{bib:BotmannNT2009, bib:HuthME2018} This may lead to new advances in one- and two-dimensional\cite{bib:LepinayNatureNanotech2017, bib:RossiNatureNanotech2017} magnetic force microscopy \cite{bib:RossiNL2019} and magnetic resonance force microscopy.\cite{bib:RugarNature2004, bib:FischerNJP2019, bib:NicholPRX2013} Our technique may also be employed with semiconducting nanowire resonators made from e.g. GaN, SiC, and InAs\cite{bib:HenryNL2007, bib:NiguesNComm2015, bib:PairisPRL2019, bib:BraakmanNT2019} as well as microfabricated top-down resonators.\cite{bib:YeoNN2014, bib:DefoortPRL2014, bib:MathewNL2015, bib:GilSantosNN2015, bib:HeritierNL2018, bib:GhadimiScience2018, bib:RoyScience2018, bib:PaulitschkeNL2019}


This work is supported by the ERC advanced Grant 692876, the Foundation Cellex, the CERCA Programme, AGAUR, Severo Ochoa (SEV-2015-0522), the Grants FIS2015-69831-P, MAT2017-82970-C2-1-R, and MAT2017-82970-C2-2-R of MINECO, the Fondo Europeo de Desarrollo Regional (FEDER), and the project E13\_17R from Aragon Regional Government (Construyendo Europa desde Aragón). This project has received funding from the European Union's Horizon 2020 research and innovation programme under the Marie Sk\l{}odowska-Curie grant agreement No 665884. P. Verlot acknowledges support from the ERC starting grant 758794 'Q-ROOT'. 

Experimental help by L. Casado from the Laboratory of Advanced Microscopies (LMA) is acknowledged.

\begin{suppinfo}

\section{Non-linearities observed via e-beam electromecnanical coupling}\label{nonlin}

\noindent Figure \ref{fig:nonlin} shows how the spectrum determined by e-beam electromechanical coupling evolves during the deposition of Pt on a carbon nanotube. For this measurement, spectra were recorded every $\approx0.32$\,s with a resolution bandwidth $BW\approx5$ kHz over a frequency span of $\approx400$\,kHz. At each time step, multiple equidistantly-spaced peaks can be observed (labeled $n=1,2,3,...10$), which correspond to harmonics of the same fundamental mode. This clearly points out that the resonator is deep in the non-linear regime of the detection.\cite{bib:TsioutsiosNL2017} Furthermore, the graph shows an increase in the signal-to-noise ratio and peak intensities as the particle grows. This indicates an increase of the coupling strength of the e-beam electromechanical coupling due to a larger volume interacting with the e-beam and a higher secondary electron (SE) yield. Furthermore, it hints towards the presence of self-oscillations.\cite{bib:TsioutsiosNL2017}

\begin{figure*}[hb]
\includegraphics[width=.5\linewidth]{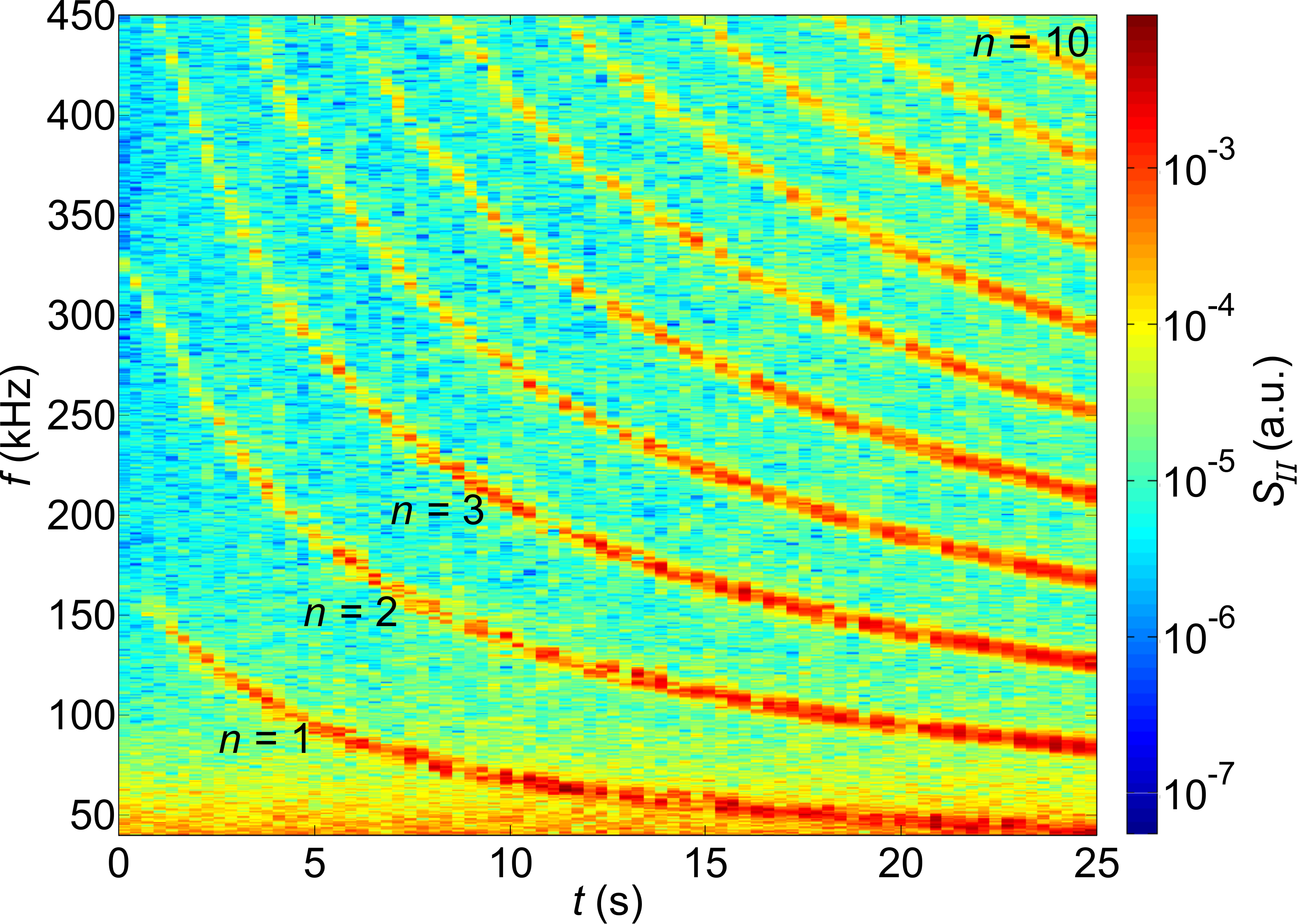}
\caption{Evolution of the spectra recorded via e-beam electromechanical coupling during the deposition of Pt on a carbon nanotube.}
\label{fig:nonlin}
\end{figure*}

\section{Optomechanical measurement of $f_\textrm{res}$}

\noindent In order to confirm the resonance frequency $f_\textrm{res}$ determined by e-beam electromechanical coupling, we conducted optomechanical measurements.\cite{bib:TavernarakisNC2018} The measurement in Fig. \ref{fig:opto} shows the thermally-driven peak with the resonance frequency $f_\textrm{res}=57.04\,\textrm{kHz}$ at room temperature. This value is reasonably close to the one observed using the e-beam electromechanical coupling $f_\textrm{res}=56.1\,\textrm{kHz}$.
\begin{figure*}[hb]
\includegraphics[width=.5\linewidth]{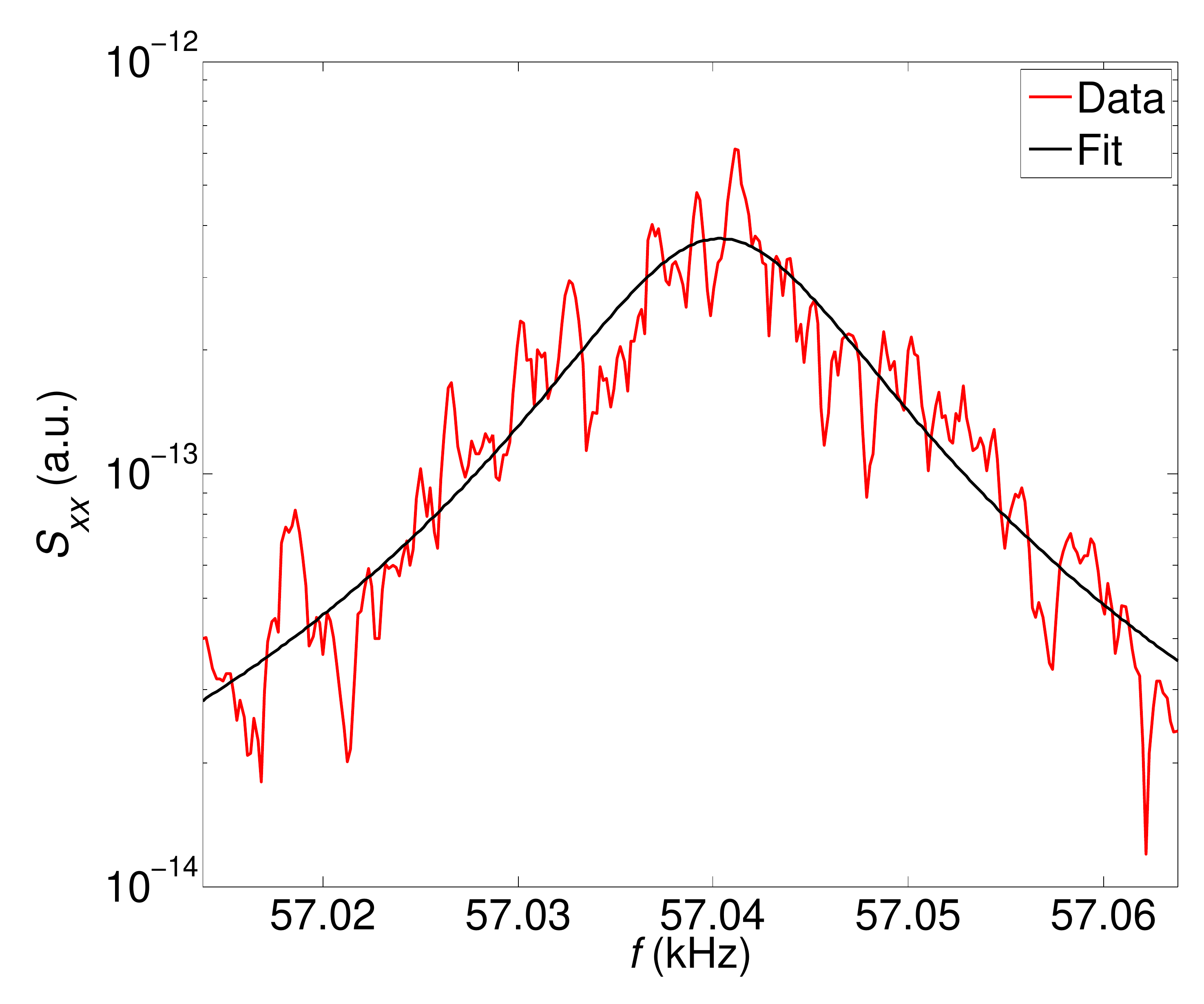}
\caption{Optomechanical measurement of the thermally-driven resonance peak (red) of the nanotube shown in Fig. 1 of the main text at room temperature and Lorentzian fit (black).}
\label{fig:opto}
\end{figure*}

\end{suppinfo}

\providecommand{\latin}[1]{#1}
\makeatletter
\providecommand{\doi}
  {\begingroup\let\do\@makeother\dospecials
  \catcode`\{=1 \catcode`\}=2 \doi@aux}
\providecommand{\doi@aux}[1]{\endgroup\texttt{#1}}
\makeatother
\providecommand*\mcitethebibliography{\thebibliography}
\csname @ifundefined\endcsname{endmcitethebibliography}
  {\let\endmcitethebibliography\endthebibliography}{}

\end{document}